\documentclass[12pt]{iopart} 
\usepackage{bm}
\usepackage{amsfonts,amssymb}
\usepackage{epsfig}

\begin{document}

\title[Control of chaos in Hamiltonian systems]{Controlling chaotic transport in a Hamiltonian model of interest to magnetized plasmas}
\author{G. Ciraolo}
\address{Facolt\`a di Ingegneria, Universit\`a di Firenze,
via S. Marta, I-50129 Firenze, Italy, and I.N.F.M. UdR Firenze}
\author{C. Chandre, R. Lima, M. Vittot}
\address{CPT-CNRS, Luminy Case 907, F-13288 Marseille Cedex 9,
France}

\author{M. Pettini}
\address{Istituto Nazionale di Astrofisica, Osservatorio
Astrofisico di Arcetri, Largo Enrico Fermi 5, I-50125 Firenze, Italy,
I.N.F.M. UdR Firenze and I.N.F.N Sezione di Firenze}

\author{C. Figarella, P. Ghendrih}
\address{Association Euratom-CEA, DRFC/DSM/CEA, CEA Cadarache,
F-13108 St. Paul-lez-Durance Cedex, France}
\eads{\mailto{ciraolo@arcetri.astro.it}, \mailto{chandre@cpt.univ-mrs.fr}, \mailto{lima@cpt.univ-mrs.fr}, \mailto{vittot@cpt.univ-mrs.fr}, \mailto{pettini@arcetri.astro.it}, \mailto{charles.figarella@cea.fr}, \mailto{philippe.ghendrih@cea.fr}}

\begin{abstract}
We present a technique to control chaos in Hamiltonian systems which are close to integrable.
By adding a small and simple {\em control term} to the perturbation, the system becomes more regular than the original one. We
apply this technique to a model that reproduces turbulent ${\bf E}\times {\bf B}$ drift and show numerically that the control is able to drastically reduce chaotic transport. 
\end{abstract}

\submitto{\JPA}
\maketitle

\section{Introduction}
In this article, the problem we address is how to control chaos in 
Hamiltonian systems which are close to integrable.  
We consider the class of Hamiltonian systems that can be written
in the form $H=H_0+\epsilon V$ that is an integrable 
Hamiltonian $H_0$ (with action-angle variables)
plus a small perturbation $\epsilon V$.
\\ \indent The problem of control
in Hamiltonian systems is the following one: For the perturbed Hamiltonian
$H_0+\epsilon V$, the aim is to devise a control term $f$ such that
the dynamics of the controlled Hamiltonian $H_0+\epsilon V+f$
has more regular trajectories (e.g.~on invariant tori) or less diffusion
than the uncontrolled one. In practice, we do not require that the controlled Hamiltonian is integrable, but only that it has a more regular behavior than the original system. This allows us to tailor the control term following specific requirements.\\
Obviously $f=-\epsilon V$ is a solution since the resulting Hamiltonian is integrable. However, it is a useless solution
since the control is of the same order as the perturbation.  \\

 For practical purposes, the desired control term should be small
(with respect to the perturbation $\epsilon V$), localized in phase space
(meaning that the subset of phase space where $f$ is non-zero is finite
or small enough),
or $f$ should be of a specific shape (e.g. a sum of given Fourier modes, or with a certain regularity). Moreover, the control should be as simple as possible in order to be implemented in experiments. 
Therefore, the control appears to be a trade-off between the requirement on the reduction of chaos and the requirement on the specific shape of the control.\\
\\ \indent In this article, we apply the method of control based on Ref.~\cite{michel} and developed in Refs.~\cite{guido1,guido2}. We implement numerically an algorithm for finding a control term $f$
of order $O(\epsilon^2)$ such that $H=H_0+\epsilon V+ f$
is integrable. This control term is expressed as a series whose terms can be explicitly and easily computed by recursion. This approach of control is ``dual'' with respect to KAM theory~: KAM theory is looking at coordinates making the system integrable. In this method of control, we slightly modify the Hamiltonian such that the resulting controlled system is integrable.  
It is shown on an example of particles in a turbulent ${\bf E}\times {\bf B}$ field that truncations of this control term $f$ provide effective control terms that significantly reduce chaotic transport. We show that with an electric potential with several frequencies, the partial control potential is time-dependent (contrary to the one used in Refs.~\cite{guido1,guido2} and that it is able to drastically reduce the diffusion of test particle trajectories. \\

\section{Control theory of Hamiltonian systems.}
\label{sec:2}

In this section, we follow the exposition of control theory developed in Ref.~\cite{michel}. Let $\mathcal A$ be the Lie algebra of real functions of class $C^\infty$ defined on
phase space. For $H\in \mathcal A$, let $\{H\}$ be the linear operator
acting on $\mathcal A$  such that
$$
\{H\}H^{\prime}=\{H,H^{\prime}\},
$$
for any $H^{\prime}\in \mathcal A$, where $\{\cdot~,\cdot\}$ is the Poisson bracket. The time-evolution of a function $V\in {\mathcal A}$ following the flow of $H$ is given by
$$
\frac{dV}{dt}=\{ H\} V,
$$  
which is formally solved as
$$
V(t)=e^{t\{H\}}V(0),
$$
if $H$ is time independent, and where
$$
e^{t\{H\}}=\sum_{n=0}^{\infty}\frac{t^n}{n!}\{H\}^n.
$$
Any element $V\in{\mathcal A}$ 
such that \( \{{H}\}{V} =0 \), is constant under the flow of \( {H} \),
i.e.
$$
\forall t\in {\mathbb R}, \qquad e^{{t} \{{H}\}} {V} = {V}.
$$
Let us now fix a Hamiltonian $H_0\in{\mathcal A}$.
The vector space \( \mathrm{Ker} \{{H_0}\} \) is the set of constants
of motion and it is a sub-Lie algebra of $\mathcal A$. The operator
\(\{{H_0}\} \) is not invertible since a derivation has always 
a non-trivial kernel. For
instance \( \{ {H_0} \} ({H_0}^\alpha) = 0 \) for any $\alpha$ such
that \( {H_0}^\alpha \in {\mathcal A} \). 
Hence we consider a pseudo-inverse of \( \{{H_0}\} \).
We define a linear operator $\Gamma$ on $\mathcal A$ such that
\begin{equation}
\{{H_0}\}^{2}\ \Gamma = \{{H_0}\},
\label{gamma}
\end{equation}
i.e.
$$
\forall V\in {\mathcal A}, \qquad \{H_0,\{H_0,\Gamma V\}\}=\{H_0,V\}.
$$
If the operator $\Gamma$ exists, it is not unique in general. Any other choice
$\Gamma^{\prime}$ satisfies that the range $\rm{Rg}(\Gamma^{\prime}-\Gamma)$ is included into the kernel $ \rm{Ker}(\{H_0\}^2)$.
\\ \indent We define the {\em non-resonant} operator $\mathcal N$ and the 
{\em resonant} operator $\mathcal R$ as
\begin{eqnarray*}
&& {\mathcal N} = \{H_0\}\Gamma,\\
&& {\mathcal R} = 1-{\mathcal N},
\end{eqnarray*}
where the operator $1$ is the identity in the
algebra of linear operators acting on \( \mathcal {A} \). 
We notice that Eq.~(\ref{gamma}) becomes
$$
\{{H_0}\} \mathcal R = 0,
$$
which means that \( {\rm Rg} \mathcal R \) is
included into \( {\rm Ker} \{{H_0}\} \).
A consequence
is that any element ${\mathcal R} V$ is constant under the flow of $H_0$, i.e. 
$e^{t\{H_0\}}{\mathcal R}V={\mathcal R}V$. We notice that when $\{H_0\}$
and $\Gamma$ commute, $\mathcal R$ and $\mathcal N$ are projectors, i.e.
$\mathcal R^2=\mathcal R$ and $\mathcal N^2=\mathcal N$. Moreover, in this case we have ${\rm Rg} \mathcal R = {\rm Ker} \{{H_0}\} $, i.e.\ the constants of motion are the elements $\mathcal{R}V$ where $V\in \mathcal{A}$.
\\ \indent Let us now assume that $H_0$ is integrable with action-angle variables 
$({\bf A},\bm{\varphi})\in B\times \mathbb{T}^n $ where $B$ is an open set
of ${\mathbb R}^n$ and ${\mathbb T}^n$ is the $n$-dimensional torus, so that
$H_0=H_0({\bf A})$ and the Poisson bracket $\{H,H^{\prime}\}$ between two Hamiltonians is 
$$
\{H,H^{\prime}\}=\frac{\partial H}{\partial{\bf A}}\cdot
\frac{\partial H^{\prime}}{\partial{\bm{\varphi}}}-
\frac{\partial H}{\partial{\bm{\varphi}}}\cdot
\frac{\partial H^{\prime}}{\partial{\bf A}}.
$$
The operator $\{H_0\}$ acts on $V$ given by
$$
V=\sum_{{\bf k}\in \mathbb Z^n}V_{\bf k}({\bf A})e^{i{\bf k}\cdot{\bm\varphi}},
$$
as
$$
\{H_0\}V({\bf A},\bm{\varphi})=\sum_{\bf k}i{\bm \omega}({\bf A})\cdot{\bf k}~V_{\bf k}({\bf A})e^{i{\bf k}\cdot\bm\varphi},
$$
where the frequency vector is given by
$$
{\bm \omega}({\bf A})=\frac{\partial H_0}{\partial{\bf A}}. 
$$
A possible choice of $\Gamma$ is
$$
\Gamma V({\bf A},\bm{\varphi})=
\sum_{{\bf k}\in{\mathbb Z^n}\atop{\omega({\bf A})\cdot{\bf k}\neq0}}
\frac{V_{\bf k}({\bf A})}
{i{\bm \omega}({\bf A}) \cdot{\bf k}}~~e^{i{\bf k}\cdot{\bm\varphi}}.
$$
We notice that this choice of $\Gamma$ commutes with $\{H_0\}$.
\\ \indent For a given $V\in{\mathcal A}$, ${\mathcal R} V$ is the resonant 
part of $V$ and ${\mathcal N} V$ is the non-resonant part:
\begin{eqnarray}
&&{\mathcal R}V=\sum_{\bf k~}
V_{\bf k}({\bf A})\chi(\bm\omega({\bf A})\cdot{\bf k}=0)e^{i{\bf k}\cdot{\bm\varphi}},\label{eqn:RV}\\ 
&&{\mathcal N}V=\sum_{\bf k~}
V_{\bf k}({\bf A})\chi(\bm\omega({\bf A})\cdot{\bf k}\neq0)e^{i{\bf k}\cdot{\bm\varphi}},
\end{eqnarray}
where $\chi(\alpha)$ vanishes  when $\alpha$ is wrong and it is equal to $1$
when  $\alpha$ is true.\\

From these operators defined for the integrable part $H_0$, we construct a control term for the perturbed Hamiltonian $H_0+V$ where $V\in {\mathcal A}$, i.e.\ we construct $f$ such that $H_0+V+f$
is canonically conjugate to $H_0+\mathcal R V$.\\

\noindent {\em Proposition 1: } ~~For $V \in {\mathcal A}$ and $\Gamma$ constructed
from $H_0$, we have the following equation
\begin{equation}
e^{\{\Gamma V\}}(H_0+V+f)=H_0+{\mathcal R} V,
\label{prop1}
\end{equation}
where
\begin{equation}
f(V)=e^{-\{\Gamma V\}}{\mathcal R}V+\frac{1-e^{-\{\Gamma
V\}}}{\{\Gamma V\}} {\mathcal N} V -V.
\end{equation}

We notice that the operator $(1-e^{-\{\Gamma V\}})/\{\Gamma V\}$
is well defined by the expansion
$$
\frac{1-e^{-\{\Gamma V\}}}{\{\Gamma V\}}=
\sum_{n=0}^{\infty}\frac{(-1)^n}{(n+1)!}\{\Gamma V\}^n.
$$
We can expand the control term in power series as
\begin{equation}
f(V)=\sum_{n=1}^{\infty}\frac{(-1)^n}{(n+1)!}\{\Gamma V\}^n
(n{\mathcal R}+1)V.
\label{expansion_f}
\end{equation}
We notice that
if $V$ is of order $\epsilon$, $f(V)$ is of order $\epsilon^2$.\\
{\em Proof}:
Since $e^{\{\Gamma V\}}$ is invertible, Eq.~(\ref{prop1}) gives
$$
f(V)=(e^{-\{\Gamma V\}}-1)H_0+e^{-\{\Gamma V\}}{\mathcal R}V-V.
$$
We notice that the operator $e^{-\{\Gamma V\}}-1$ can be divided by 
$\{\Gamma V\}$
$$
f(V)=\frac{e^{-\{\Gamma V\}}-1}{\{\Gamma V\}}\{\Gamma V\}H_0+
e^{-\{\Gamma V\}}{\mathcal R}V-V.
$$
By using the relations
$$
\{\Gamma V\}H_0=\{\Gamma V,H_0\}=-\{H_0\}\Gamma V,
$$
and
$$
\{H_0\}\Gamma={\mathcal N},
$$
we have
$$
f(V)=e^{-\{\Gamma V\}}{\mathcal R}V+
\frac{1-e^{-\{\Gamma V\}}}{\{\Gamma V\}}{\mathcal N}V-V.
~~\Box
$$
This result can also be obtained by using a perturbation series.\\
\indent Proposition 1 tells that the addition of a well chosen 
small control term $f$ makes the Hamiltonian canonically
conjugate to $H_0+{\mathcal R} V$. 

\noindent {\em Proposition 2 :} ~~The flow of $H_0+V+f$ is conjugate to the flow of $H_0+{\mathcal R}V$:
$$
\forall {t} \in {\mathbb R} , \qquad e^{{t} \{{H_0} + {V} + {f}\}} =
e^{-\{\Gamma {V}\}} ~ e^{{t} \{{H_0}\}}~ e^{{t} \{\mathcal R {V}\}}
~ e^{\{\Gamma {V}\}}. 
$$

The remarkable fact is that
the flow of ${\mathcal R}V$ commutes with the one of $H_0$, since
$\{H_0\}{\mathcal R} = 0$. This allows the splitting of the flow of 
$H_0 +{\mathcal R} V$ into a product. \\
 
We recall that
$H_0$ is {\em non-resonant} iff
$$
\forall {\bf A}\in B, ~~  \chi \left(\omega({\bf A})\cdot{\bf k}=0\right)=\chi ( {\bf k=0} ).
$$ 
If $H_0$ is non-resonant then with the addition of a
control term $f$, the Hamiltonian $H_0+V+f$ is 
canonically conjugate to the integrable Hamiltonian $H_0+{\mathcal R} V$
since ${\mathcal R} V$ is only a function of the 
actions [see Eq.~(\ref{eqn:RV})].
\\ \indent If $H_0$ is resonant and ${\mathcal R} V=0$, the controlled
Hamiltonian $H=H_0+V+f$ is conjugate to $H_0$.
\\ In the case ${\mathcal R} V=0$, the series (\ref{expansion_f})
which gives the expansion of the control term $f$,
can be written as
\begin{equation}
f(V)=\sum_{s=2}^{\infty}f_s,
\label{exp_f_rv_0}
\end{equation}
where $f_s$ is of order $\epsilon^s$ and given by the
recursion formula
\begin{equation}
f_s=-\frac{1}{s}\{\Gamma V,f_{s-1}\},
\label{recursion}
\end{equation}
where $f_1=V$.

\noindent {\em Remark :} A similar approach of control has been developed by G. Gallavotti in Refs.~\cite{gallavotti2,gallavotti1,gentile}. The idea is to find a control term (named {\it counter term}) only depending on the actions, i.e.\ to find $N$ such that
$$
H({\bf A},{\bm \varphi})=H_0({\bf A})+V({\bf A},{\bm \varphi})-N({\bf A}),
$$
is integrable. For isochronous systems, that is
$$
H_0({\bf A})={\bm \omega}\cdot {\bf A},
$$
or any function $h({\bm \omega}\cdot {\bf A})$,
it is shown that if the frequency vector satisfies a Diophantine condition and if the perturbation is sufficiently small and smooth, such a control term exists, and that an algorithm to compute it by recursion is provided by the proof. We notice that the resulting control term $N$ is of the same order as the perturbation, and has the following expansion
$$
N({\bf A})={\mathcal R} V+\frac{1}{2} {\mathcal R} \{\Gamma V \} V+O(\varepsilon^3),
$$
where we have seen from Eq.~(\ref{eqn:RV}) that ${\mathcal R}V$ is only a function of the actions in the non-resonant case where ${\bm \omega}$ is non-resonant which is a crucial hypothesis in Gallavotti's renormalization approach. Otherwise, a counter-term which only depends on the actions ${\bf A}$ cannot be found. In what follows, we will see that the integrable part of the Hamiltonian is always resonant for the cases we consider, namely for the ${\bf E}\times{\bf B}$ drift motion in the guiding center approximation.

\section{Application to a model of ${\bf E}\times {\bf B}$ drift motion}

In the guiding center approximation, the equations of motion
of a charged particle in presence of a strong toroidal magnetic field
and of a nonstationary electric field are~\cite{northrop}
\begin{equation}
{\dot{\bf x}}= \frac{d}{dt}{x \choose y}=\frac{c}{B^2}{\bf E}({\bf
x},t)\times {\bf B}= \frac{c}{B}{-\partial_y V (x,y,t)\choose
\partial_x V (x,y,t)},
\label{guidcent}
\end{equation}
where $V$ is the electric potential, ${\bf E}=-{\bf\nabla} V$,
and ${\bf B}=B {\bf e}_z$. 
The spatial coordinates $x$ and $y$ where $(x,y)\in \mathbb{R}^2$ play the role of the canonically
conjugate variables and the electric potential $V(x,y,t)$
is the Hamiltonian of the problem.
To define a model we choose
\begin{equation}
V ({\bf x},t)=\sum_{{\bf n}\in{\bf{\mathbb Z^2}}}{V_{\bf n}}\sin \left [ \frac{2\pi}{L}{\bf n}\cdot {\bf x}
+\varphi_{\bf n}-\omega ({\bf n})t\right ],
\end{equation}
where $\varphi_{\bf n}$ are random phases and $V_{\bf n}$ decrease as
a given function of $\vert {\bf n}\vert$, in agreement with experimental data
\cite{anormal_exp}.
In principle, one should use for $\omega ({\bf n})$ the dispersion
relation for electrostatic drift waves (which are thought to be
responsible for the observed turbulence) with a frequency broadening
for each ${\bf n}$ in order to model the experimentally observed
spectrum~\cite{anormal_exp} $S({\bf n},\omega)$.

By rescaling space and time, we can always assume that $L=2\pi$.
In this article, we use a simplified potential. We keep the random phases in order to model a turbulent electric potential and we use a simplified broadening of the spectrum for each ${\bf n}$. We consider the following electric potential~:
\begin{equation}
\label{eqn:Vqp}
V(x,y,t)=\sum_{k=1}^K\sum_{m,n=1}^N \frac{a_k}{(n^2+m^2)^{3/2}} \sin ( nx+my+\varphi_{kmn}-\omega_k t ),
\end{equation}
where $\varphi_{kmn}$ are random phases.
Here we consider a quasiperiodic approximation of the turbulent electric potential with a finite number $K$ of frequencies. We assume that $\omega_k\not= 0$ (see remark below).\\
First we map this Hamiltonian system with $1+K/2$ degrees of freedom to an autonomous Hamiltonian with $1+K$ degrees of freedom by considering that $\theta_k=\omega_k t \mbox{ mod} 2\pi$ are angles. We denote $E_k$ the action conjugate to $\theta_k$. This autonomous Hamiltonian is
\begin{equation}
H(x,y,{\bm \theta},{\bf E})={\bm \omega} \cdot {\bf E}+ \sum_{k,m,n}\frac{a_k}{(n^2+m^2)^{3/2}}  \sin (nx+my+\varphi_{kmn}-\theta_k).
\end{equation}
The integrable part of the Hamiltonian from which the operators $\Gamma$, ${\mathcal R}$ and ${\mathcal N}$ are constructed is isochronous
$$
H_0={\bm \omega} \cdot {\bf E}.
$$
We notice that $H_0$ is resonant (since it does not depend on the action variable $x$). Also, we notice that the frequency vector ${\bm \omega}$ is in general resonant since the potential contains also harmonics of the main frequencies.\\
The action of $\Gamma$, ${\mathcal R}$ and ${\mathcal N}$ on functions $W$ of the form 
$$
W(x,y,{\bm \theta})=\sum_{\bm\nu} W_{\bm \nu}(x,y)e^{i {\bm \nu}\cdot {\bm \theta}},
$$
is
\begin{eqnarray*}
&& \Gamma W=\sum_{\bm\nu} \frac{W_{\bm \nu}(x,y)}{i {\bm \omega}\cdot {\bm \nu}}\chi ({\bm \omega}\cdot {\bm \nu}\not= 0) e^{i {\bm \nu}\cdot {\bm \theta}},\\
&& {\mathcal R} W=\sum_{\bm\nu} W_{\bm \nu}(x,y) \chi ({\bm \omega}\cdot {\bm \nu}= 0) e^{i {\bm \nu}\cdot {\bm \theta}},\\
&& {\mathcal N} W=\sum_{\bm\nu} W_{\bm \nu}(x,y) \chi ({\bm \omega}\cdot {\bm \nu}\not= 0) e^{i {\bm \nu}\cdot {\bm \theta}}.
\end{eqnarray*}
The action of $\Gamma$ and ${\mathcal R}$ on the potential $V$ given by Eq.~(\ref{eqn:Vqp}) is 
\begin{eqnarray*}
&& \Gamma V=\sum_{k,m,n} \frac{a_{k}}{\omega_k(n^2+m^2)^{3/2}} \cos (nx+my+\varphi_{kmn}-\theta_k ),\\
&& {\mathcal R} V=0.
\end{eqnarray*}
From the action of these operators, we compute the control term using Eqs.~(\ref{exp_f_rv_0}) and (\ref{recursion}). For instance, the expression of $f_2$ is
\begin{eqnarray}
\lefteqn{ f_2(x,y,t)= \frac{1}{4}\sum_{k,m,n\atop k',n',m'} \frac{a_k a_{k'} (n'm-nm')}{\omega_k (n^2+m^2)^{3/2}(n'^2+m'^2)^{3/2}} \times} \nonumber \\
& & \quad \left( \sin [(n+n')x+(m+m')y+\varphi_{kmn}+\varphi_{k'm'n'}-(\omega_k+\omega_{k'})t] \right. \nonumber  \\
& & \left. +\sin [(n-n')x+(m-m')y+\varphi_{kmn}-\varphi_{k'm'n'}-(\omega_k-\omega_{k'})t] \right). \label{Vqpf2}
\end{eqnarray}

\begin{figure}
\begin{center}
\epsfig{file=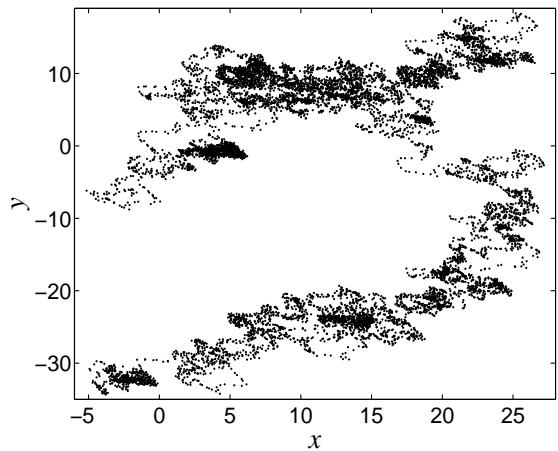,width=7.5cm,height=6cm}
\end{center}
\caption{Poincar\'e surface of section of a trajectory obtained for
Hamiltonian (\ref{eqn:Vqp}) assuming $a=0.8$.}
\label{figure1}
\end{figure}

\begin{figure}
\begin{center}
\epsfig{file=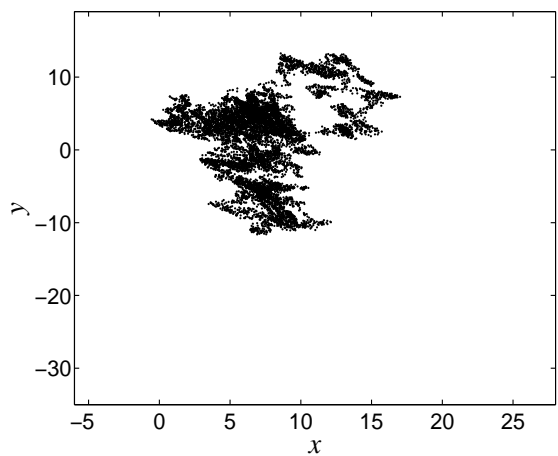,width=7.5cm,height=6cm}
\end{center}
\caption{Poincar\'e surface of section of a trajectory obtained for
the same initial condition as in Fig.~\ref{figure1} and adding the control term 
(\ref{Vqpf2}) to Hamiltonian (\ref{eqn:Vqp}).}
\label{figure2}
\end{figure}

\begin{figure}[ht]
\begin{center}
\epsfig{file=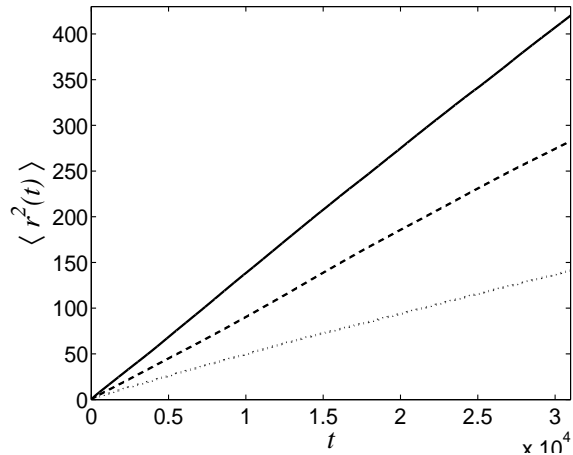,width=7.5cm,height=6cm}
\end{center}
\caption{Mean square displacement $\langle r^2 (t)\rangle $ versus
time $t$ obtained for  Hamiltonian
(\ref{eqn:Vqp}) with the control term 
(\ref{Vqpf2}) for three different values of $a=0.7$ (dotted line), $a=0.8$ (dashed line), $a=0.9$ (continuous line).}
\label{figure3}
\end{figure}

\begin{figure}[ht]
\begin{center}
\epsfig{file=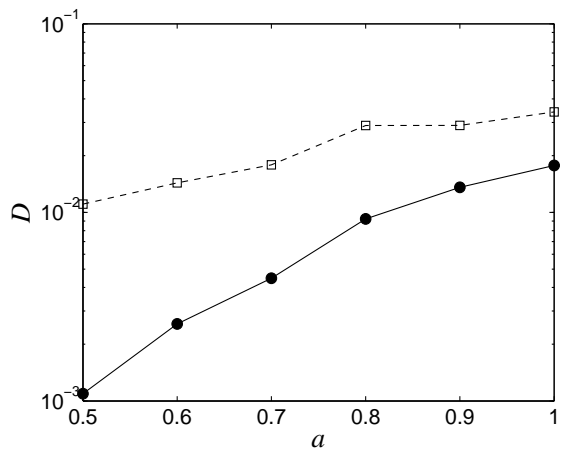,width=7.5cm,height=6cm}
\end{center}
\caption{Diffusion coefficient $D$ versus $a$
obtained for Hamiltonian (\ref{eqn:Vqp}) (open squares) and Hamiltonian
(\ref{eqn:Vqp}) plus control term (\ref{Vqpf2}) (full
circles).}
\label{figure4}
\end{figure}

{\em Remark:} Similar calculations can be done in the case where there is a zero frequency, e.g., $\omega_0=0$ and $\omega_k\not= 0$ for $k\not= 0$. The first term of the control term is 
$$
f_2=-\frac{1}{2} \left\{\Gamma V, ({\mathcal R} +1) V\right\}, 
$$
where
$$
{\mathcal R}V=\sum_{m,n} \frac{a_0}{(n^2+m^2)^{3/2}} \sin (nx+my+\varphi_{0mn}).
$$
We implement numerically the control term by computing test particle trajectories in the original electric potential and the controlled potential obtained by adding $f_2$ to $V$. Even if Proposition 1 tells that the addition of the control term $f$ makes the dynamics integrable, the numerical simulations are used to test the efficiency of some truncations of this control term which provide more tractable control potentials.\\
In Refs.~\cite{guido1,guido2}, the electric potential was chosen with only one frequency. It was shown that truncations of the control term are able to drastically reduce chaotic transport of charged test particles in this potential. It was also shown that approximations of this control term are still able to reduce chaos and provide simpler control potentials. In particular, it is possible to reduce the amplitude of the control term $f_2$ which means that one can inject less energy and still get sufficient control. Also it is possible to truncate the Fourier series of the control term and keeping only the main Fourier components of the control in order to get sufficient stabilization.
We perform numerical experiments for an electric potential that contains two frequencies. For instance, we use two frequencies $\omega_1=1$ and $\omega_2=\sqrt{2}$ and $a_1=a$ and $a_2=a/2$ where $a$ ranges from 0.5 to 1.
With the aid 
of numerical simulations (see Ref.~\cite{marc88}
for more details on the numerics), we check the effectiveness of
the control by comparing the diffusion properties of 
particle trajectories obtained from Hamiltonian (\ref{eqn:Vqp}) and
from the same Hamiltonian with the control term (\ref{Vqpf2}).
Figures~\ref{figure1} and \ref{figure2} show the Poincar\'e surfaces of section of two
typical trajectories (issued from the same initial conditions) computed without
and with the control term respectively. Similar pictures are obtained 
for many other randomly chosen initial conditions.
A clear evidence is found for a
relevant reduction of the diffusion in presence of the 
control term (\ref{Vqpf2}).\\ \indent In order 
to study the diffusion properties of the system, we have
considered a set of $\mathcal M$ particles (of order $100$) 
uniformly distributed at
random in the domain $0\leq x,y\leq 1$ at $t=0$. We have computed the
mean square displacement $\langle r^2 (t) \rangle$ as a function of
time
$$
\langle r^2 (t) \rangle = \frac{1}{\mathcal M} \sum_{i=1}^{\mathcal M}
{|{\bf x}_i(t) - {\bf x}_i(0)|}^2
$$
where ${\bf x}_i(t),~i=1,\dots,\mathcal M$ is the position of the
$i$-th particle at time $t$ obtained by integrating Hamilton's equations 
with initial conditions ${\bf x}_i(0)$.
Figure \ref{figure3} shows $\langle r^2 (t) \rangle$ for three different values of
$a$.
For the range of parameters we consider, the behavior 
of $\langle r^2 (t)\rangle$ is always found to be 
linear in time for $t$ large enough. The
corresponding diffusion coefficient is defined as
\[
D= \lim_{t \rightarrow\infty}{{\langle r^2 (t) \rangle} \over t}~.
\]
Figure \ref{figure4} shows the values of $D$ as a function of $a$ with and without 
control term. It clearly shows a significant decrease of the diffusion 
coefficient when the control term is added.
As expected, the action of the control term gets weaker as $a$ is
increased towards the strongly chaotic phase. Moreover, we notice that for $a>1.3$ the diffusion is larger with control than without. 

\ack
 This work is part of an ongoing collaboration between the University of Florence, the Arcetri Astrophysical Observatory (INAF), the Center for Theoretical Physics (Marseille) and the Department of Research on the Controlled Fusion (CEA Cadarache). We acknowledge the financial support from the Association Euratom/CEA (contract V 3382.001). We are grateful to G.\ Gallavotti and J. Laskar for fruitful discussions.

\end{document}